\documentclass[journal]{IEEEtran}

\usepackage{cite}
\usepackage{amsmath,amssymb,amsfonts}
\usepackage{algorithmic}
\usepackage{graphicx}
\usepackage{textcomp}
\usepackage{booktabs}
\usepackage{float}
\usepackage{comment}
\usepackage{xcolor}
\usepackage{tabularx}
\usepackage[square,sort,comma,numbers]{natbib}
\usepackage{tikz}
\usetikzlibrary{shapes.geometric, arrows}

\tikzstyle{startstop} = [rectangle, rounded corners, minimum width=2cm, minimum height=1cm,text centered, draw=black]
\tikzstyle{io} = [trapezium, trapezium left angle=70, trapezium right angle=110, minimum width=2cm, minimum height=1cm, text centered, draw=black]
\tikzstyle{process} = [rectangle, minimum width=2cm, minimum height=1cm, text centered, text width=2cm, draw=black]
\tikzstyle{decision} = [diamond, minimum width=2cm, minimum height=1cm, text centered, draw=black]
\tikzstyle{arrow} = [thick,->,>=stealth]

\begin{document}

\title{Blackbox Small-Signal Modeling of Grid-Connected Inverters in Asymmetrical Power Grids}

\author{Airan~Frances,~\IEEEmembership{Member,~IEEE,}
        Dionisio~Ramirez,~\IEEEmembership{Senior~Member,~IEEE,}
        and~Javier~Uceda,~\IEEEmembership{Life~Fellow,~IEEE}
\thanks{\copyright{} 2023 IEEE. Personal use of this material is permitted. Permission from IEEE must be obtained for all other uses, in any current or future media, including reprinting/republishing this material for advertising or promotional purposes, creating new collective works, for resale or redistribution to servers or lists, or reuse of any copyrighted component of this work in other works. DOI: 10.1109/TPEL.2023.3296786}%
\thanks{The authors are with the Centro de Electr{\'o}nica Industrial (CEI), Universidad Polit{\'e}cnica de Madrid, 28006, Madrid, Spain, e-mail: (see http://www.cei.upm.es/).}
\thanks{This manuscript is an extended version of the conference paper: A. Franc{\'e}s, D. Ram{\'\i}rez and J. Uceda, "Blackbox Small-signal Modeling of Grid-Connected Inverters under Unbalanced Conditions," 2022 IEEE 13th International Symposium on Power Electronics for Distributed Generation Systems (PEDG), Kiel, Germany, 2022, pp. 1-6.}}

\markboth{IEEE Transactions on Power Electronics,~Vol.~38, No.~10, October~2023}%
{Frances \MakeLowercase{\textit{et al.}}: Blackbox Small-Signal Modeling of Three-Phase Inverters under Unbalanced Conditions}

\maketitle

\begin{abstract}
Power electronic converters are envisaged to be key enablers of modern electric power distribution systems. Grid-connected three-phase inverters are widely used in Smart Grids and microgrids, but also in standard grids. They provide controllability and dynamic decoupling capabilities, which are fundamental in the integration of renewable sources and storage systems and in specialized applications such as FACTS. Nevertheless, power electronics-based systems can exhibit dynamic interactions, which may lead to power quality issues. Although the electrical model of each element of these systems is important for their system-level analysis, they are rarely available. Blackbox modeling strategies are useful for obtaining behavioral models of commercial electronic converters. Most blackbox modeling strategies are focused on dc-dc electronic converters; however, some works have extended these concepts to three-phase converters by means of the dq framework, under the assumption of symmetrical and balanced conditions. This work proposes a new structure in the sequence domain to represent the dynamic behavior of grid-connected commercial converters in asymmetrical conditions. Experimental tests have been performed on a three-phase inverter to identify its blackbox model and its performance has been validated during an asymmetric voltage sag. 
\end{abstract}

\begin{IEEEkeywords}
Asymmetrical conditions, blackbox, grid-side converter, modeling, system identification, three-phase inverter, unbalanced grid.
\end{IEEEkeywords}

\section{Introduction}\label{sec_Intr}
\IEEEPARstart{P}{ower} Electronic Converters (PECs) are expected to become ubiquitous in power distribution systems \cite{future_boro}. PECs play a very important role in various strategic sectors such as the energy sector, enabling a high integration of renewable sources and storage elements and supporting the smart grid concept \cite{Energy_storage_power_electronics_Molina_2017,PEC_enabling_Oliver_Cian_2012}; or the transportation sector, replacing mechanical, hydraulic, or pneumatic systems in more electric aircrafts \cite{More_Electric_aircraft_Sarlioglu_Morris_2015}, all-electric ships \cite{All-electric_ships_Sulligoi_Menis_2016} or electric vehicles \cite{Power_Electronics_Electric_vehicles_Emadi_2008,Vehicle_home_oportunities_Liu_Gao_2013}.

System identification techniques have been widely used to represent the dynamic behavior of PECs for different purposes \cite{System_identification_Ljung_1999}, such as control design, stability analyzes, system-level integration, filter and protection system designs, among others. When applied to PECs, blackbox models refer to the use of system identification techniques to obtain electrical models able to represent the dynamic behavior of devices about which the available information is scarce. 

Most of the blackbox model structures are focused on dc-dc PECs, as the powerful tools related with the linear theory can be easily applied to this kind of systems. These structures can be classified according to the kind of response they are able to reproduce. Linear structures \cite{unterminated_arnedo,bb_valdivia,unterminated_Cvetkovic_TPE_2013} are useful, due to their simplicity, and are very suitable for linear systems or nonlinear systems under small-signal conditions. Static nonlinear structures \cite{Nonlinear_modeling_Hammerstein_Alonge_Tumminaro_2007,HW_oliver,hybrid_cvet} have been proposed to capture nonlinearities in the steady-state response, while the dynamic response is considered linear, which are very useful to include the effect of protection systems or linear systems with nonlinear control references. Dynamic nonlinear structures \cite{poly_arnedo, Frances2019_APEC_PVTFM,Frances_19_Access} have been proposed for systems with nonlinear responses working in variable environments such as dc microgrids \cite{Frances2016_APEC} or electric ships \cite{Frances_18_TII}. A comparison among the different blackbox models structures for dc-dc PECs can be found in \cite{Frances2017_TSG}.

Some of these structures have been extended to ac systems, either in single-phase or three-phase systems. The idea is to use the rotating d-q coordinates to transform ac variables in dc variables, which is necessary to represent the dynamic response of these systems by means of transfer functions. In \cite{Unterminated_three_phase_Cvetkovic_2011, Naziris_2020_access} the linear structure is proposed for three-phase systems, whereas in \cite{BB_Three_phase_VSI_Valdivia_2010} the different structures are studied for the case of a voltage source inverter. In \cite{Guarderas_2019}, the nonlinear structure is implemented for a Grid-Connected Inverter (GCI) and in \cite{Naziris_2019} for battery chargers for electric vehicles. Recently, a number of studies have been presented related to the online identification of three-phase PEC and grid impedances in the $dq$-frame \cite{dq_impedance_measurement_Xiongfei_TPEL_2021,dq_grid_impedance_identification_Roinila_JESTPE_2022}. Some of the challenges are to use a single perturbation signal to identify the impedances to avoid shifts in the operating point during the identification process, or the coupling effect between converter and grid impedances. These models are very interesting for designing adaptive controllers and for performing online impedance-based stability analysis \cite{dq_impedance_stability_Roinila_TPEL_2018}.

These extensions are valid under the assumption of symmetric and balanced conditions.  From an analytical approach, the effect of asymmetrical grids on the dynamic behavior of the converters and its stability has been studied \cite{control_stability_asymmetrical_grid_Guerrero_TPEL2022}, and the importance of accounting for the coupling effects between positive and negative sequences has been demonstrated in \cite{couplings_sequences_stability_Xiongfei_Blaabjerg_TPEL2016}. However, from a blackbox perspective, this problem has not been considered. This paper proposes a blackbox small-signal structure able to represent GCIs under asymmetrical conditions. The idea is based on the Fortescue's theorem to represent the unbalanced phasor as the sum of symmetrical balanced phasors that can be represented as dc signals using the Park transformation. This work is an extended version of the conference paper \cite{Frances_PEDG_2022}. The resulting model can be used to analyze the dynamic performance of commercial-off-the-shelf inverters when connected in series or parallel with other PECs or when interfaced with any grid impedance. As part of its small-signal model is the output admittance of the inverter, this information could be used to perform impedance-based stability analysis considering asymmetrical grid impedances, which can also be represented in the $dq$ sequence domain \cite{Frances_APEC_2023_dq_unbalanced_impedances}, using the generalized Nyquist criterion \cite{GNC_stability_1977}.

The paper is organized as follows. In Section~\ref{SII} the small-signal structure able to represent three-phase inverters under asymmetrical conditions is described. In Section \ref{SIII} the proposed blackbox modeling strategy is identified and validated. Finally, in Section \ref{Conclusions} the conclusions are exposed.

\section{Small-Signal Structure of Three-Phase Inverters under Asymmetrical Conditions} \label{SII}
The first step in the definition of the small-signal model of a PEC is to identify and sort the variables as inputs or outputs. In general, the system variables are the electrical signals of the input and output ports, where the externally imposed variables (voltage or current) will be set as input and the remaining variable (current or voltage) will be set as output of the model. In the case of controlled PEC, the control references can be considered as additional inputs. Finally, in ac systems, the frequency can also be included as an input in case it is variable and it influences the dynamic behavior of the PEC.

\begin{figure}
	\centering
	\includegraphics[scale=0.95]{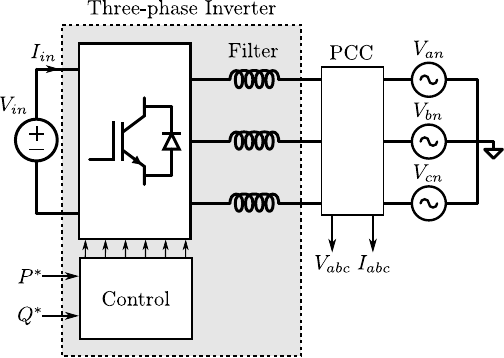}
	\caption{Three-phase grid-feeding inverter scheme.}
	\label{grid_feeding_scheme}
\end{figure}

In this paper, the small-signal structure is particularized for a GCI working as a grid-feeding PEC \cite{Control_PEC_AC_microgrids_Rocabert_Rodriguez_2012} connected to an asymmetrical power grid, as depicted in Fig.~\ref{grid_feeding_scheme}. Consequently, the input variables are the imposed input and output voltages, whereas the input and output currents are the outputs of the model. Besides, the active and reactive power references of the control can be set as inputs of the model. In this case, the grid frequency does not have a relevant influence on the dynamic response of the $d$ and $q$ components of the current, hence it has not been included as an input. However, in cases where droop control is implemented, the frequency has a strong impact on the output current and it should be considered as input of the model.

\begin{figure}
\centerline{\includegraphics[width=84mm]{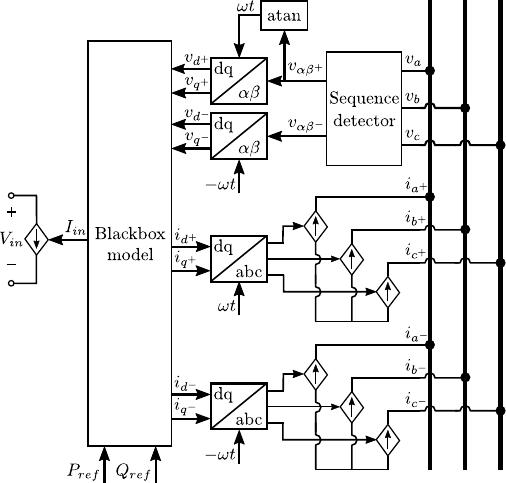}}
\caption{Blackbox model interface between the dc and the three-phase ac ports using the $d$ and $q$ symmetrical components.}
\label{bb_model_interface}
\end{figure}

\begin{figure}
	\centering
	\includegraphics[scale=0.95]{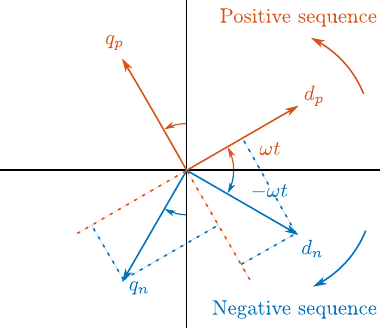}
	\caption{D-Q framework representation of the positive and negative sequences.}
	\label{dq_symmetric}
\end{figure}

In Fig.~\ref{bb_model_interface} the interface between the model and the dc and ac ports is depicted. First, the three-phase ac signals are divided into their positive and negative sequences. Notice that as the system has three wires, the zero sequence has been neglected in this analysis. Second, the positive sequence is used for synchronization. Third, the positive and negative sequences are represented in their respective synchronous frames, i.e. the positive sequence is synchronized using $\omega t$ (counterclockwise rotation), whereas the negative sequence uses $-\omega t$ (clockwise rotation); see Fig.~\ref{dq_symmetric}. Notice that if only the counterclockwise rotating framework is used for both the positive and negative sequences, the negative sequence would appear as a second harmonic. Using the clockwise rotating framework, the negative sequence can also be represented with constant $d$ and $q$ components.

The different parts of the model structure are described in the following sections.


\subsection{Sequence detector}

 According to the Fortescue's theorem, a three-wire three-phase unbalanced signal ($v_{abc}$) can be expressed as the sum of two balanced sequences, the positive ($v^+_{abc}$) and negative components ($v^-_{abc}$):

\begin{equation}
    v_{abc}=v^+_{abc}+v^-_{abc},
\end{equation}

\noindent where the positive and negative sequences can be obtained by means of the Lyon transformation \cite{Teodorescu_Book_2010}:

\begin{equation} \label{v+}
    v^+_{abc}=[T^+]v_{abc}= \dfrac{1}{3}
\left[ \begin{array}{ccc}
1 & a & a^2 \\
a^2 & 1 & a \\
a & a^2 & 1
\end{array} \right]
\left[ \begin{array}{c}
v_a \\
v_b \\
v_c 
\end{array} \right]
\end{equation}

\begin{equation} \label{v-}
    v^-_{abc}=[T^-]v_{abc}= \dfrac{1}{3}
\left[ \begin{array}{ccc}
1 & a^2 & a \\
a & 1 & a^2 \\
a^2 & a & 1
\end{array} \right]
\left[ \begin{array}{c}
v_a \\
v_b \\
v_c 
\end{array} \right],
\end{equation}

\noindent and $a$ is a 120$^\circ$ phase-shift operator, which can be defined as follows:

\begin{equation}
    a=e^{\frac{2\pi}{3}i},
\end{equation}

\noindent where $i$ is the imaginary unit. In the $\alpha \beta$ plane, (\ref{v+}) and (\ref{v-}) can be expressed as:

\begin{equation}
    v^+_{\alpha \beta}=[T_{\alpha \beta}][T^+][T_{\alpha \beta}]^{-1}v_{\alpha \beta}
\end{equation}

\begin{equation}
    v^-_{\alpha \beta}=[T_{\alpha \beta}][T^-][T_{\alpha \beta}]^{-1}v_{\alpha \beta},
\end{equation}

\noindent where $[T_{\alpha \beta}]$ is the Clarke transformation given by:

\begin{equation} \label{Clarke}
    [T_{\alpha \beta}]= \dfrac{2}{3} \left[ \begin{array}{ccc}
1 & -\dfrac{1}{2} & -\dfrac{1}{2} \\ \\
0 & \dfrac{\sqrt{3}}{2} & - \dfrac{\sqrt{3}}{2} 
\end{array} \right].
\end{equation}

The result of these transformations is the following relationship:

\begin{equation} \label{Positive sequence}
    v^+_{\alpha \beta}=\dfrac{1}{2} \left[ \begin{array}{cc}
1 & -q  \\
q & 1
\end{array} \right] v_{\alpha \beta}
\end{equation}

\begin{equation} \label{Negative sequence}
    v^-_{\alpha \beta}=\dfrac{1}{2} \left[ \begin{array}{cc}
1 & q  \\
-q & 1
\end{array} \right] v_{\alpha \beta},
\end{equation}

\noindent where $q$ is a -90$^\circ$ phase shift (in-quadrature) operator, which can be defined as:

\begin{equation}
    q=e^{-\frac{\pi}{2}i}.
\end{equation}

\begin{figure}
\centerline{\includegraphics[width=80mm]{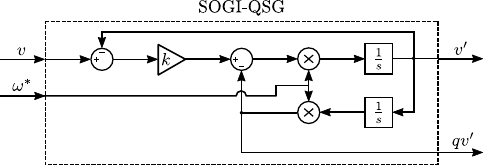}}
\caption{SOGI-QSG block diagram.}
\label{SOGI_QSG}
\end{figure}

There are several ways to implement a -90$^\circ$ phase shift to a sinusoidal signal. The most straightforward method is to use a transport delay of a quarter of a cycle. However, for system identification purposes, this strategy has the drawback of providing discontinuous and erroneous results during more than 2 cycles. A more sophisticated approach is the Second Order Generalized Integrator in-Quadrature Signal Generator (SOGI-QSG), which is able to produce a continuous response and to provide the correct result in about 1 cycle \cite{Teodorescu_Book_2010}. In Fig.~\ref{SOGI_QSG} the block diagram of the SOGI-QSG is depicted, where $\omega^*$ is the fundamental frequency of the sinusoidal input signal $v$ and the transfer functions from the input ($v$) to the outputs ($v'$ and $qv'$) are:

\begin{equation} \label{v'}
    \dfrac{v'}{v}(s)=\dfrac{k\omega^*s}{s^2+k\omega^*s+\omega^{*^2}}
\end{equation}

\begin{equation} \label{qv'}
    \dfrac{qv'}{v}(s)=\dfrac{k\omega^{*^2}}{s^2+k\omega^*s+\omega^{*^2}}.
\end{equation}

\begin{figure}
\centerline{\includegraphics[width=80mm]{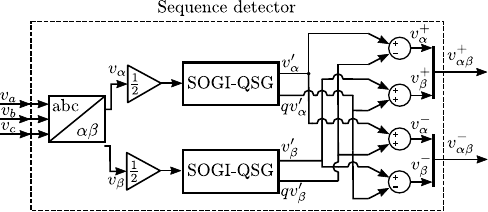}}
\caption{Sequence detector block diagram.}
\label{Sequence detector}
\end{figure}

Finally, the sequence detector consists of a Clarke transformation (\ref{Clarke}) and the implementation of (\ref{Positive sequence}) and (\ref{Negative sequence}) by means of the SOGI-QSG depicted in Fig.~\ref{SOGI_QSG}. The block diagram of the sequence detector is represented in Fig.~\ref{Sequence detector}.

It is important to notice that the inputs of the blackbox model will be affected by the dynamic of the SOGI-QSG, presented in (\ref{v'}) and (\ref{qv'}). Practically, this fact imposes a limitation on the maximum frequency at which the model can be estimated, which is related to the minimum time in which the sequence detector provides the correct result after a perturbation. Using a faster sequence detector or taking into account the dynamic imposed by the SOGI-QSG would be some key points to improve the capabilities of the proposed model. In cases where the frequency is considered as an input of the model, a DSOGI-FLL \cite{Teodorescu_Book_2010} can be implemented to obtain the grid frequency and used as an input to the SOGI-QSG. Similarly, the dynamic response of the DSOGI-FLL will influence the range of frequencies at which the dynamic response of the PEC can be estimated.

\subsection{Synchronization}

A synchronization system is needed in the model structure in order to obtain the $d$ and $q$ components of the positive and negative sequences of the signals. Assuming that the signals are not distorted or that the harmonic content has been properly filtered, the arc-tangent function is the best option for this application due to its instantaneous response, where the phase angle is defined as:

\begin{equation}
    \theta = arctan \left( \dfrac{v^+_\beta}{v^+_\alpha} \right).
\end{equation}

Notice that, since the synchronization is performed using the positive sequence component of the voltage (see Fig.~\ref{bb_model_interface}), the SOGI-QSG block has filtered most of the harmonic content of the signals, according to (\ref{v'}) and (\ref{qv'}), which mitigate the drawbacks of using the arc-tangent function as synchronization system.

\subsection{Small-signal model}

The small-signal model comprises a collection of transfer functions that describe the dynamic response of every output when each of the inputs is perturbed. Consequently, the number of transfer functions is equal to the number of input variables by the number of outputs.

As synchronization systems usually make the $q$ component of the positive sequence of the voltage, $v_{q^+}$, equal to zero, this input is not considered. For the sake of simplicity and without loss of generality, $V_{in}$ and $Q_{ref}$ will also be considered constant and not included as inputs of the model. 

Finally, splitting the ac signals into the $d$ and $q$ components of their positive and negative sequences, the small-signal model can be described as (\ref{SS_Model}), where "$\land$" represents small-signal variables and ``+'' and ``-'' refer to the positive and negative sequences, respectively. The names of the transfer functions have a two-character sub index, where the first character refers to the input and the second to the output. For instance, $Y_{d^-q^- }$ refers to the transfer function between the $d$ component of the negative sequence of the three-phase voltage and the $q$ component of the negative sequence of the three-phase current.

\begin{figure}
\centerline{\includegraphics[width=80mm]{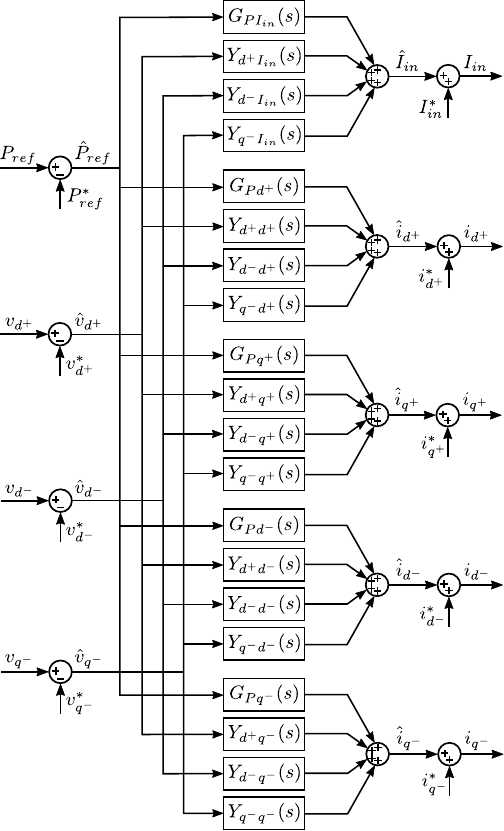}}
\caption{Block diagram of the blackbox small-signal model.}
\label{SSM_block}
\end{figure}

In Fig.~\ref{SSM_block} the block diagram of the small-signal model is represented according to (\ref{SS_Model}), where ``*'' refers to the operating point where the model is identified. Each of the transfer functions of the model is identified from the response of each of the outputs when each input is perturbed,  holding all other inputs
constant.

\begin{figure}
\centerline{\includegraphics[width=85mm]{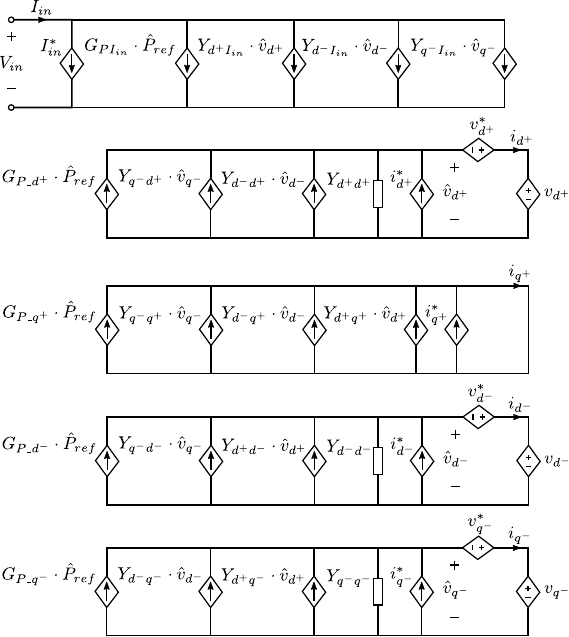}}
\caption{Blackbox small-signal electrical equivalent model.}
\label{electrical_equivalent}
\end{figure}

\begin{multline} \label{SS_Model} 
\left( \begin{matrix}
\hat i_{in} \\
\hat i_{d}^+ \\
\hat i_{q}^+ \\
\hat i_{d}^- \\
\hat i_{q}^-
\end{matrix} \right) = \left( \begin{matrix}
G_{PI_{in}} (s) & Y_{d^+I_{in}} (s) \\
G_{Pd^+} (s) & Y_{d^+d^+} (s) \\
G_{Pq^+} (s) & Y_{d^+q^+} (s) \\
G_{Pd^-} (s) & Y_{d^+d^-} (s) \\  
G_{Pq^-} (s) & Y_{d^+q^-} (s) \\ \end{matrix}\right. \\
\left. \begin{matrix}
Y_{d^-I_{in}} (s) & Y_{q^-I_{in}} (s) \\
Y_{d^-d^+} (s) & Y_{q^-d^+} (s) \\
Y_{d^-q^+} (s) & Y_{q^-q^+} (s) \\
Y_{d^-d^-} (s) & Y_{q^-d^-} (s) \\
Y_{d^-q^-} (s) & Y_{q^-q^-} (s) \\
\end{matrix} \right) \left( \begin{matrix}
\hat P_{ref} \\
\hat v_{d}^+ \\
\hat v_{d}^-  \\
\hat v_{q}^-
\end{matrix} \right).
\end{multline}

\subsubsection{DC operating point}

The DC operating point of all the variables involved in the small-signal model is constant and corresponds to the operating point in which the transfer functions were identified. However, the input current is a special case as, in spite of being a dc variable, it can show oscillations in the steady state, due to the presence of negative sequence in the three-phase port. To include this effect in the small-signal model, the input current reference, $I^*_{in}$, will be defined using a power balance between the input power and the active power of the three-phase port:

\begin{equation} \label{Iin_OP}
    I^*_{in}=\dfrac{P}{V_{in}\cdot\eta},
\end{equation}

\noindent where $V_{in}$ is the input voltage, $\eta$ is the efficiency of the converters, and $P$ is the active power of the three-phase port. The active power can be expressed as a function of the $d$ and $q$ components of the three-phase voltages and currents:

\begin{equation} \label{eq:Pdq}
    P=\dfrac{3}{2} \left( v_d \cdot i_d + v_q \cdot i_q\right),
\end{equation}

\noindent and the $d$ and $q$ components of the three-phase voltages can be split into its positive and negative components (the current components are analogous):

\begin{equation} \label{eq:vdpn}
    v_d=v^p_{dp}+v^p_{dn}
\end{equation}
\begin{equation} \label{eq:vqpn}
    v_q=v^p_{qp}+v^p_{qn},
\end{equation}

\noindent where the super index $p$ refers to the positive framework, and the sub-index has two characters: the first refers to the component ($d$ or $q$) and the second indicates the sequence ($p$ for positive and $n$ for negative). Furthermore, to use only dc variables, the $d$ and $q$ components of the negative sequence in the positive reference can be expressed as a function of the $d$ and $q$ components of the negative sequence in the negative reference (which are constant). This relationship can be obtained using the trigonometric projections of the $d$ and $q$ axes of the negative sequence on the $d$ and $q$ axes of the positive sequence (see Fig.~\ref{dq_symmetric}):

\begin{equation} \label{eq:pdn}
    v^p_{dn}=v^n_{dn} \cdot cos(2\omega t) +v^n_{qn} \cdot sin(2\omega t)
\end{equation}
\begin{equation} \label{eq:pqn}
    v^p_{qn}=-v^n_{dn} \cdot sin(2\omega t) +v^n_{qn} \cdot cos(2\omega t).
\end{equation}

Finally, using (\ref{eq:vdpn}), (\ref{eq:vqpn}), (\ref{eq:pdn}) and (\ref{eq:pqn}), in (\ref{eq:Pdq}), the active power an be expressed as a function of the $d$ and $q$ components of the symmetrical components in their respective frameworks:

\begin{equation} \label{eq:Pdq}
    P= P_a + P_c \cdot cos (2\omega t) + P_s \cdot sin(2\omega t), 
\end{equation}

\noindent where $P_a$ is the average value of the active power, and $P_c$ and $P_s$ are second-harmonic oscillations of the active power. These terms are defined as follows:

\begin{equation} \label{eq:Pa}
    P_a=\dfrac{3}{2} \left( v^p_{dp} \cdot i^p_{dp} + v^p_{qp} \cdot i^p_{qp} + v^n_{dn} \cdot i^n_{dn} + v^n_{qn} \cdot i^n_{qn}\right)
\end{equation}
\begin{equation} \label{eq:Pc}
    P_c= \dfrac{3}{2} \left(v^p_{dp} \cdot i^n_{dn} + v^p_{qp} \cdot i^n_{qn} + v^n_{dn} \cdot i^p_{dp} + v^n_{qn} \cdot i^p_{qp}\right)
\end{equation}
\begin{equation} \label{eq:Ps}
    P_s= \dfrac{3}{2} \left(v^p_{dp} \cdot i^n_{qn} - v^p_{qp} \cdot i^n_{dn} - v^n_{dn} \cdot i^p_{qp} + v^n_{qn} \cdot i^p_{dp}\right).
\end{equation}

\section{Experimental Tests} \label{SIII}
\begin{figure}
\centerline{\includegraphics[width=85mm]{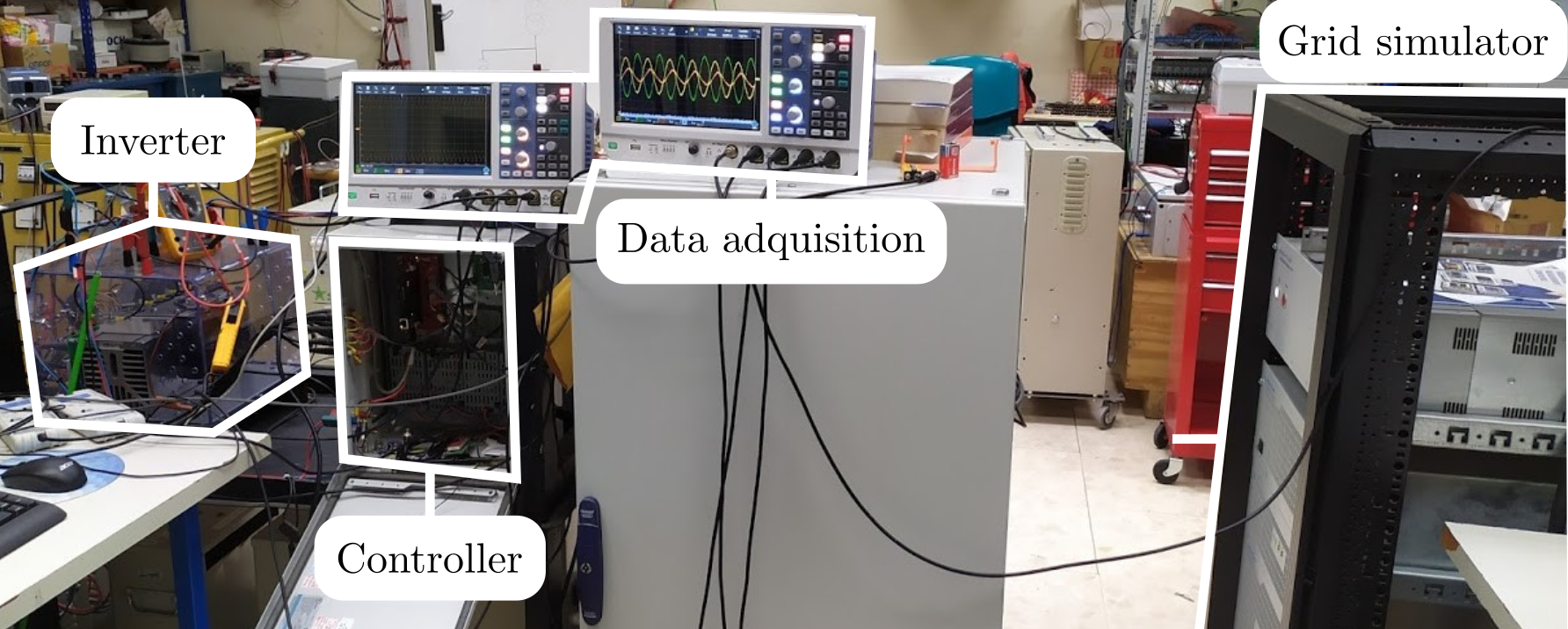}}
\caption{Experimental setup used to perform the identification and validation tests.}
\label{Experimental_Setup}
\end{figure}

A Semikron Semiteach VSC controlled by a dual-core F28M35H52C Concerto microcontroller has been used for the experimental tests, Fig.~\ref{Experimental_Setup}. The control structure consists of a decoupled current control, in which the reference is set as a function of the active and reactive power reference to be injected into the grid. The control includes a sequence detector in which the in-quadrature signals are obtained by means of a quarter of a cycle delay. Then, the positive sequence of the current is the one compared with the current reference, and the positive sequence of the voltage is the one fed forward to the control. The synchronization is made by applying the arc-tangent function to the $\alpha\beta$ components of the positive sequence of the voltage. Notice that the focus of this work is not on the design of the converter and its control loop, but on the validation of the proposed blackbox model architecture to describe the small-signal dynamic behavior of the system under asymmetric conditions. It is also worth noticing that different control decisions, such as the use or not of a sequence detector, the use of the positive sequence of the current and/or the voltage, the synchronization method, the design of the PI regulators, the decoupling terms using nominal values or measured values, among other details, would impact the system behavior in asymmetric conditions.

A four-quadrant REGATRON TC.ACS has been employed to create the specific perturbations in the $d$ and $q$ components of the positive and negative sequences of the grid voltage necessary to identify the transfer functions of the proposed model. In general, the output impedance of the source converter can interact with the output impedance of the converter under test and influence the measurements. However, in this case, as the bandwidth of the REGATRON TC.ACS is much higher than the bandwidth of the converter, this effect can be neglected in the range of frequencies of interest. To validate the identified model, an asymmetric voltage sag was imposed to generate an asymmetrical and unbalanced scenario. 

\subsection{Model identification} 

The identification process consists of applying the appropriate tests to obtain each of the transfer functions described in (\ref{SS_Model}). The number of tests is equal to the number of inputs of the model, in this case, four. Applying superposition, in each test only one of the inputs is perturbed while keeping the rest of the inputs constant (its small-signal contribution equal to zero), and the corresponding column of transfer functions of (\ref{SS_Model}) is identified from the dynamic response of each of the outputs. Practically, the three-phase currents are measured and introduced into a sequence detector and an $\alpha\beta/dq$ transformation to extract the $d$ and $q$ components of its positive and negative sequences (see Fig.~\ref{bb_model_interface}). 

\begin{figure*}
\centerline{\includegraphics[width=160mm]{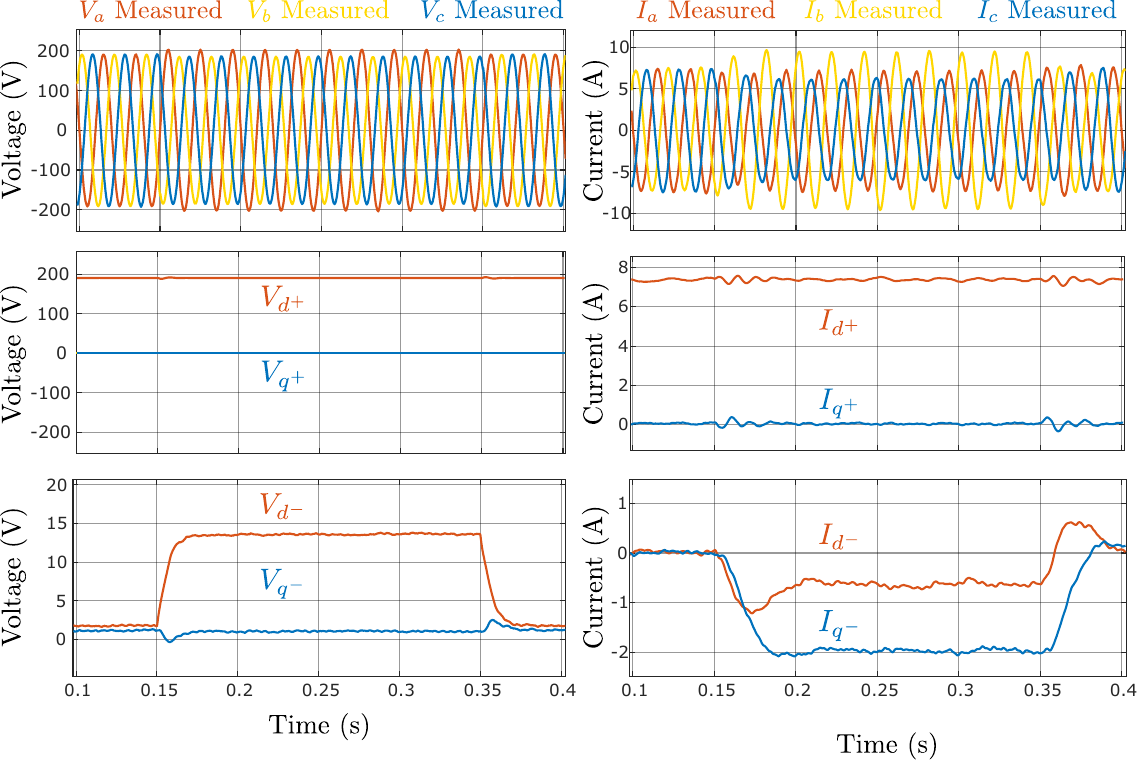}}
\caption{Third experimental test performed in the model identification process. A step in the $d$ component of the negative sequence of the three-phase voltage is injected,  holding all other inputs constant. The plots on the left side show the measured voltages and the components of the positive and negative sequences. The plots on the right side present the measured currents and the components of the positive and negative sequences.}
\label{vdn_test}
\end{figure*}

The first test consists of perturbing the reference of the active power while keeping the $d$ and $q$ components of the positive and negative sequences of the three-phase voltage constant. A step was introduced in the reference of the active power of the controller.

The second test involves the perturbation of the $d$ component of the positive sequence of the three-phase voltage, holding all other inputs constant. A balanced step was imposed in the amplitude of the three-phase voltage.

The third test implies the perturbation of the $d$ component of the negative sequence (represented in the negative reference frame) of the three-phase voltage, holding all other inputs constant. A negative sequence in phase with the positive sequence was injected. The mathematical expression of the three-phase voltage used for this test was:

\begin{align} \label{third_test}
    \begin{split}
         v_a(t)=&A_p \cdot cos(\omega t) + A_n \cdot cos(\omega t) \\
       v_b(t)=&A_p \cdot cos(\omega t-\dfrac{2\pi}{3}) + A_n \cdot cos(\omega t+\dfrac{2\pi}{3}) \\
       v_c(t)=&A_p \cdot cos(\omega t+\dfrac{2\pi}{3}) + A_n \cdot cos(\omega t-\dfrac{2\pi}{3}),
    \end{split}
\end{align}

\noindent where $A_p$ and $A_n$ are the amplitudes of the positive and negative sequence, respectively.
      
Finally, the fourth test comprises the perturbation of the $q$ component of the negative sequence of the three-phase voltage, holding all other inputs constant. A negative sequence with a 90 degrees phase shift with respect to the positive sequence was injected. The mathematical expression of the three-phase voltage used for this test was:

\begin{align} \label{fourth_test}
    \begin{split}
         v_a(t)=&A_p \cdot cos(\omega t) + A_n \cdot cos(\omega t-\dfrac{\pi}{2}) \\
       v_b(t)=&A_p \cdot cos(\omega t-\dfrac{2\pi}{3}) + A_n \cdot cos(\omega t+\dfrac{2\pi}{3}-\dfrac{\pi}{2}) \\
       v_c(t)=&A_p \cdot cos(\omega t+\dfrac{2\pi}{3}) + A_n \cdot cos(\omega t-\dfrac{2\pi}{3}-\dfrac{\pi}{2}).
    \end{split}
\end{align}

To illustrate the identification process, the measured signals obtained during the third test are depicted in Fig.~\ref{vdn_test}, where the plots on the left side show the measured voltages and the plots on the right side include the measured currents. Horizontally, the top plots depict the three-phase measured signals, the plots in the middle represent the components of the positive sequence, and the plots at the bottom show the components of the negative sequence. At time 0.15~s, a step in $V_{d^-}$ is injected, according to (\ref{third_test}), and at time 0.35 the perturbation is eliminated. Regarding the components of the sequences of the voltage, which are the inputs of the model, it can be seen that only $V_{d^-}$ is perturbed, while the rest are kept constant. 

\begin{figure}
\centerline{\includegraphics[width=75mm]{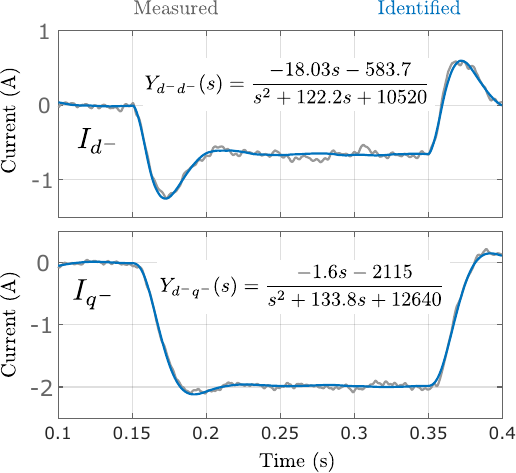}}
\caption{Comparison between the measured dynamic response and the output of the identified transfer functions for the components of the negative sequence of the current during the third test. The expressions of the identified transfer functions are displayed in their respective plots.}
\label{identified_TFs}
\end{figure}

Concerning the current, the three-phase currents become unbalanced due to the injection of negative sequence in the voltage. Regarding the components of the sequences of the current, which are the outputs of the model, it can be seen how the effect of the injection of $V_{d^-}$ has a negligible effect on $I_{d^+}$ and $I_{q^+}$. However, this test has a relevant effect on both $I_{d^-}$ and $I_{q^-}$. The transfer functions related to these last two outputs are $Y_{d^-d^-} (s)$ and $Y_{d^-q^-} (s)$ respectively. Using the Matlab System Identification Toolbox, introducing $V_{d^-}$ as input signal and $I_{d^-}$ and $I_{q^-}$ as output signals, these transfer functions were identified, as presented in Fig.~\ref{identified_TFs}.

\begin{figure} 
    \centering
    \scalebox{0.9}{
    \begin{tikzpicture}[node distance=1.6cm,align=center]
    
\node (start) [startstop] {Start};
\node (init) [io, below of=start] {Bestfit = 0 \\BestTF = 0 \\ Np = 1};
\node (dec_np) [decision, below of=init,yshift=-0.5cm] {Np $>$ 4?};
\node (init_nz) [process, below of=dec_np,yshift=-0.4cm] {Nz = 0};
\node (dec_nz) [decision, below of=init_nz,yshift=-0.5cm] {Nz $>$ Np?};
\node (estimate) [process, below of=dec_nz,yshift=-1.25cm] {Estimate a transfer function TF with \\ Np poles and Nz zeros};
\node (dec_fit) [decision, below of=estimate,yshift=-1.5cm] {TF fitness $>$ \\Bestfit + $\varepsilon$?};
\node (TF) [process, below of=dec_fit,yshift=-0.75cm] {BestTF=TF};
\node (Nz+) [process, right of=dec_fit,xshift=1.5cm] {Nz=Nz+1};
\node (Np+) [process, left of=init_nz,xshift=-1cm] {Np=Np+1};
\node (stop) [startstop,right of=dec_np,xshift=1.25cm] {BestTF};

\draw [arrow] (start) -- (init);
\draw [arrow] (init) -- (dec_np);
\draw [arrow] (dec_np) -- node[anchor=west] {no} (init_nz);
\draw [arrow] (init_nz) -- (dec_nz);
\draw [arrow] (dec_nz) -- node[anchor=west] {no} (estimate);
\draw [arrow] (estimate) -- (dec_fit);
\draw [arrow] (dec_fit) -- node[anchor=west] {yes}(TF);
\draw [arrow] (dec_fit) -- node[anchor=south] {no}(Nz+);
\draw [arrow] (TF) -| (Nz+);
\draw [arrow] (Nz+) |- (dec_nz);
\draw [arrow] (dec_nz) -| node[anchor=south,xshift=1cm] {yes} (Np+);
\draw [arrow] (Np+) |- (dec_np);
\draw [arrow] (dec_np) -- node[anchor=south] {yes} (stop);

\end{tikzpicture}
}
    \caption{Flowchart of transfer function identification from experimental data.}
    \label{Flowchart}
\end{figure}
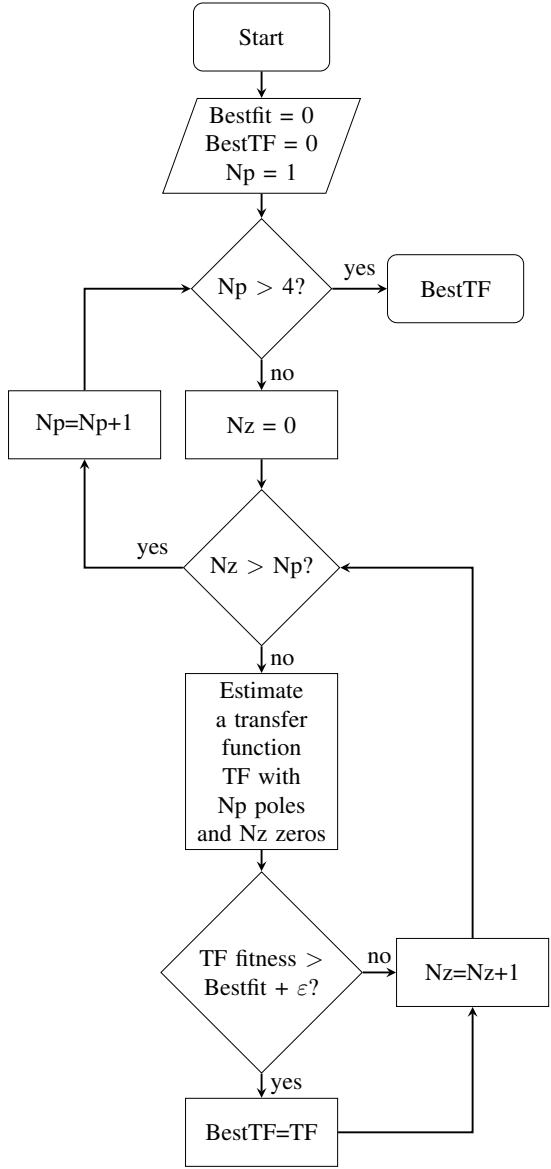

The identification of the transfer functions from the time-domain measured responses is described as follows. First, the initial average value is subtracted from all signals, such that they all start from zero. The subtracted average values of each signal define the operating point at which the identification has been performed, as detailed in Table~\ref{table:OP}. Then, the MATLAB function \textit{tfest} is implemented to identify the parameters of a transfer function of a specific number of poles a zeros that fits the measured data with better accuracy. In this work, the identification has been partially automated using the flowchart depicted in Fig.~\ref{Flowchart}. The 14 possible combinations of poles and zeros from 1 pole and 0 zeros to 4 poles and 4 zeros are tested and a tradeoff between complexity and accuracy is implemented in the selection of the best transfer function. To favor low-order transfer functions, a term $\varepsilon$ is inserted in the fitness comparison to ensure that the selected higher-order transfer functions have an improvement in accuracy of at least this amount (in this work 5\% has been chosen).

\begin{table}
    \centering
    \caption{DC Operating Point.}
    \label{table:OP}
    \centering
       \begin{tabularx}{0.95\linewidth}{m{1cm} m{5.2cm} m{1cm}}
        \toprule
        Symbol  & Description & Value \\
        \midrule
        $P^*_{ref}$  & Active power reference & 2.1 kW \\
        $v^*_{d^+}$  & $d$ component of the positive sequence of the three-phase voltage & 191 V \\
        $v^*_{d^-}$  & $d$ component of the negative sequence of the three-phase voltage & 1.8 V \\
        $v^*_{q^-}$  & $q$ component of the negative sequence of the three-phase voltage & 1.1 V \\
        $i^*_{d^+}$  & $d$ component of the positive sequence of the three-phase current & 7.4 A \\
        $i^*_{q^+}$  & $q$ component of the positive sequence of the three-phase current & 0 A \\
        $i^*_{d^-}$  & $d$ component of the negative sequence of the three-phase current & 0 A \\
         $i^*_{q^-}$  & $q$ component of the negative sequence of the three-phase current & 0 A \\
         $I^*_{in}$  & input current & 5.7 A \\
        \bottomrule
    \end{tabularx}
\end{table}

\begin{figure*}
\centerline{\includegraphics[width=160mm]{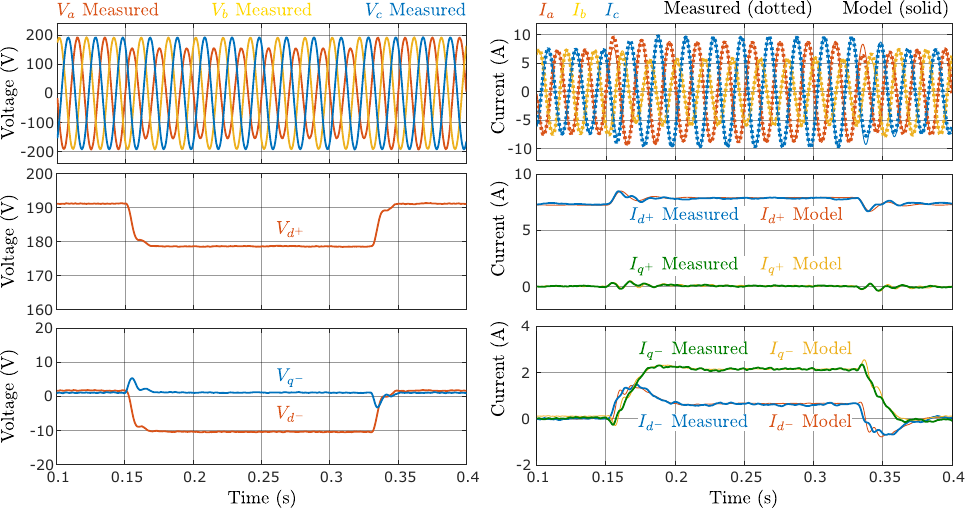}}
\caption{Experimental validation of the proposed model using a 0.2 pu voltage sag in one of the phases. The plots on the left side represent the voltage measurements. The plots on the right side present the comparison between the measured currents from the experimental setup and the blackbox model estimation.}
\label{Validation}
\end{figure*}

\begin{figure}
\centerline{\includegraphics[width=82mm]{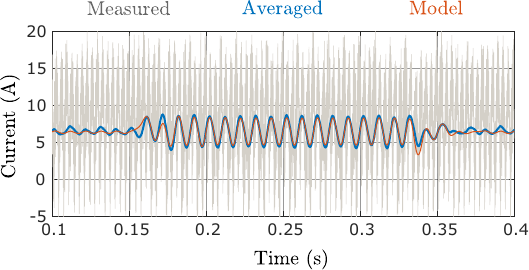}}
\caption{Comparison during a voltage sag in one of the phases among the measured input current, the averaged value of the measured input current, and the estimation of the model for the input current.}
\label{Iin}
\end{figure}

\begin{figure*}
\centerline{\includegraphics[width=160mm]{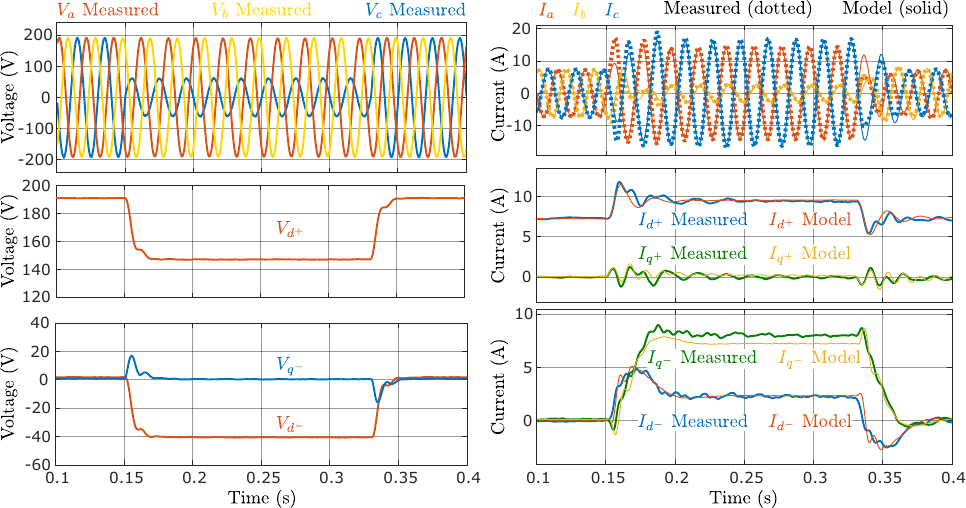}}
\caption{Experimental validation of the proposed model using a 0.7 pu voltage sag in one of the phases. The plots on the left side represent the voltage measurements. The plots on the right side present the comparison between the measured currents from the experimental setup and the blackbox model estimation.}
\label{Validation70}
\end{figure*}

\subsection{Model validation}

This test aims to validate the identified model by imposing an asymmetrical and unbalanced grid condition and comparing the response of the actual converter with the estimation of the blackbox model. The injected perturbation was a voltage sag in one of the phases, while the others remained unaffected. The amplitude of the perturbed phase changes from 190~V to 150~V, approximately, during the test, which roughly corresponds to 0.2 pu. This scenario generates changes in different components of the sequences of the voltage, which impact several components of the sequences of the current. 

In Fig.~\ref{Validation} the estimation of the model is compared with the measured signals from the experimental setup. At time $0.15~s$ the voltage sag is injected, whereas at time $0.33~s$ the voltage is restored. The organization of the signals is the same as in Fig.~\ref{vdn_test}, the plots on the left side represent voltages, whereas plots on the right side present the currents. From top to bottom, the first row shows three-phase signals and the second and the third display the components of the positive and negative sequences, respectively. 

Regarding the components of the voltage, it is observed that the injected perturbation has a remarkable effect on the $d$ components of both the positive and negative sequences, whereas the impact on the $q$ component of the negative sequence is negligible.

Concerning the current, it is noted that the effect of the perturbation is mainly reflected in the negative sequence, where both the $d$ and $q$ components are disturbed; however, there is also a small effect in the $d$ component of the positive sequence. 

The comparison between the measured current from the experimental setup and the estimation of the model demonstrates that the proposed blackbox modeling structure is able to accurately predict the static and dynamic behavior of three-phase inverters under asymmetrical and unbalanced conditions.

Regarding the input port, Fig.~\ref{Iin} shows the comparison of the measured signal and the estimation of the input current using the proposed model for the same case of a voltage sag in one of the three phases. The input current measurement (gray color) has high-frequency components which cannot be accounted for in the model, which is an averaged model. The averaged value of the input current (blue color) is also represented and it can be seen that during the voltage sag in the three-phase side, a second harmonic oscillation appears in the input current due to the effect of the negative sequence in the three-phase signals. This effect is accurately reproduced by the model (red color) by means of a power balance using (\ref{Iin_OP}) and (\ref{eq:Pdq}), where the active power of the three-phase port can be computed as a function of the rest of the dc inputs and outputs of the model, as described in (\ref{eq:Pa}), (\ref{eq:Pc}), and (\ref{eq:Ps}).

Finally,  Fig.~\ref{Validation70} shows the comparison between the estimation of the model and the measured signals from the experimental setup with a 0.7 pu voltage sag in one of the phases. The result shows that the proposed model provides a good representation of the three-phase inverter even in highly asymmetrical conditions.

\section{Conclusions} \label{Conclusions}
In the literature, blackbox models of three-phase systems assume symmetrical and balanced conditions to transform the ac signals into constant $d$ and $q$ components in the synchronous frame. This work extends the applicability of blackbox models to asymmetrical conditions considering the $d$ and $q$ components of the positive and negative sequences, each in its corresponding synchronous frame, where they remain constant even in asymmetrical scenarios. This framework is used to create a small-signal structure able to consider the dynamic relationship among the different variables of the system. The model structure is detailed, where the main elements are the sequence detector and the synchronization system. Using the arc-tangent function the synchronization is instantaneous; however, the SOGI-QSG used in the sequence detector adds some dynamic to the sequence detection, which limits the model capabilities. The specific tests necessary to identify each of the transfer functions that comprise the model have been detailed. Furthermore, a method is proposed to account for the second harmonic component that appears in the averaged response of the input current due to the presence of negative sequence in the three-phase port. Finally, the identification and validation processes are illustrated by means of experimental results.





\begin{IEEEbiography}[{\includegraphics[width=1in,height=1.25in,clip,keepaspectratio]{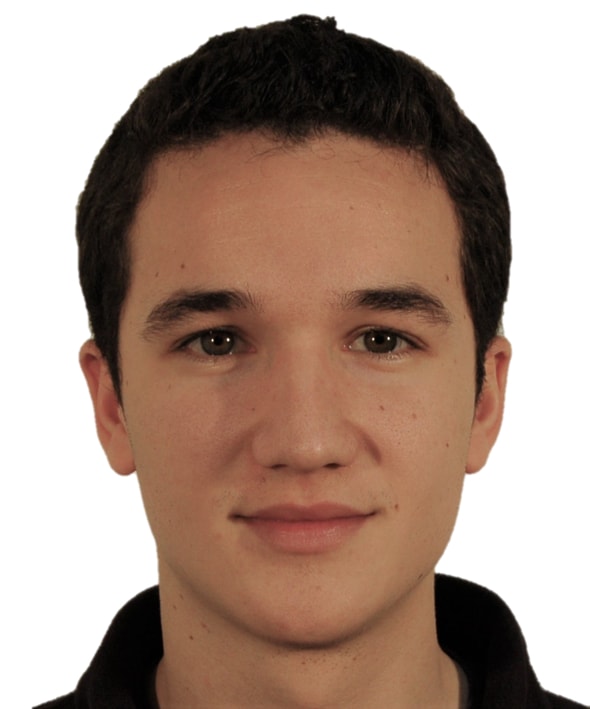}}]{Airan Frances}
(GS'16--M'18) received the M.Sc. and Ph.D. degrees in electrical engineering from the Universidad Polit{\'e}cnica de Madrid (UPM), Spain, in 2012 and 2018, respectively.

He is currently an Assistant Professor with the Department of Electrical Engineering, UPM. He participated for two years in the European Project XFEL, where he collaborated on the design
and development of dc--dc power supplies for superconducting magnets. His current research interests include modeling, control, and stability assessment of electronic power distribution systems and smart grids.
\end{IEEEbiography}

\begin{IEEEbiography}[{\includegraphics[width=1in,height=1.25in,clip,keepaspectratio]{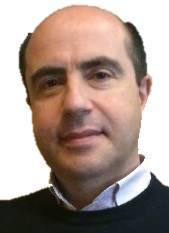}}]{Dionisio Ramirez}
(M'07--SM'15) received the M.S. degree in industrial engineering and the Ph.D. degree in electrical engineering from the Universidad
Politecnica de Madrid (UPM), Madrid, Spain, in 1997 and 2003, respectively. 

He is with the Department of Electrical Engineering, UPM. His research interests include electric control of renewable energy and DSP-based control systems.
\end{IEEEbiography}


\begin{IEEEbiography}[{\includegraphics[width=1in,height=1.25in,clip,keepaspectratio]{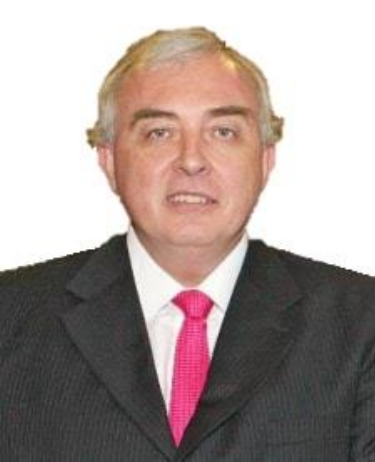}}]{Javier Uceda}
 (M'83--SM'91--F'05) received the M.Sc. and Ph.D. degrees in electrical engineering from the Universidad Polit{\'e}cnica de Madrid (UPM), Spain, in 1976 and 1979, respectively.
 
He is currently a Full Professor in electronics with the Electrical and Electronic Engineering Department, UPM. His research activity has
been developed in the field of power electronics, where he has participated in numerous national and international research projects. His research interests include switched-mode power supplies and dc--dc power converters for telecom and aerospace applications. As a result of this activity, he has published more than three hundred articles in international journals and conferences. He holds several national and international patents.

Prof. Uceda has received several individual and collective awards, such as the IEEE Third Millennium Medal, the Puig Ad{\'a}n Medal. He is also an Honorary Doctor by the Universidad Ricardo Palma in Per{\'u} and the Colegio de Posgraduados in Mexico.
\end{IEEEbiography}

\end{document}